\documentclass[%
 aip,
rsi,%
 amsmath,amssymb,
 reprint,%
]{revtex4-1}

\usepackage{graphicx}
\usepackage{dcolumn}
\usepackage{bm}
\usepackage{natbib}

\renewcommand{\eqref}[1]{Eq.~(\ref{#1})}
\newcommand{\eqsref}[1]{Eqs.~(\ref{#1})}
\newcommand{\pref}[1]{(\ref{#1})}
\newcommand{\secref}[1]{Sec.~\ref{#1}}
\newcommand{\figref}[1]{Fig.~\ref{#1}}

\newcommand{\prt}[2]{\frac{\partial #1}{\partial #2}}

\newcommand{\rd}{\mathrm{d}}
\newcommand{\ri}{\mathrm{i}}
\newcommand{\expct}[1]{\langle #1 \rangle}
\DeclareMathOperator{\im}{Im} 
\DeclareMathOperator{\re}{Re} 
\DeclareMathOperator{\Std}{Std}

\begin{document}

\preprint{AIP/123-QED}

\title{Measuring Lyapunov exponents of large chaotic systems with global coupling by time series analysis\
}

\author{Taro P. Shimizu}
 \affiliation{
 Department of Physics, Tokyo Institute of Technology, 2-12-1 Ookayama, Meguro-ku, Tokyo, 152-8551, Japan.
 }
\author{Kazumasa A. Takeuchi}%
 \affiliation{
 Department of Physics, The University of Tokyo, 7-3-1 Hongo, Bunkyo-ku, Tokyo, 113-0033, Japan.
}%
 \affiliation{
 Department of Physics, Tokyo Institute of Technology, 2-12-1 Ookayama, Meguro-ku, Tokyo, 152-8551, Japan.
 }

\date{\today}

\begin{abstract}
Despite the prominent importance of the Lyapunov exponents
 for characterizing chaos,
 it still remains a challenge to measure them for large experimental systems,
 mainly because of the lack of recurrences in time series analysis.
Here we develop a method to overcome this difficulty,
 valid for highly symmetric systems such as systems with global coupling,
 for which the dimensionality of recurrence analysis
 can be reduced drastically.
We test our method numerically with two globally coupled systems, 
 namely, logistic maps and limit-cycle oscillators with global coupling.
The evaluated exponent values are successfully compared
 with the true ones obtained by the standard numerical method.
We also describe a few techniques
 to improve the accuracy of the proposed method.
\end{abstract}

\maketitle

\begin{quotation}
Although chaos with many degrees of freedom abounds
 in a wide variety of natural systems, such as turbulence in geophysical flows
 and laboratory experiments \cite{frisch1995turbulence},
 chemical reactions \cite{Wang2000},
 and possibly cardiac arrhythmia \cite{Alonso2008},
 it still remains challenging to characterize their instability
 in a quantitative manner.
A practical method for measuring Lyapunov exponents is particularly called for,
 because the Lyapunov exponents and related concepts
 are useful to characterize various aspects of chaos,
 as well as for application purposes such as chaos control \cite{Ott-Book2002}.
In this work, we propose a method to evaluate the Lyapunov exponents
 of large chaotic systems from time series,
 which is valid for systems with a high degree of symmetry.
Focusing on globally coupled systems, and using a time series
 of a single local variable and the mean field,
 we demonstrate that our method can indeed estimate
 the full spectrum of the Lyapunov exponents correctly.
We expect that the presented idea can also be extended
 to other types of symmetric systems, paving the way toward
 experimental investigations of instability of large chaotic systems
 in the future.
\end{quotation}
\section{\label{sec:introduction}Introduction}

Instability is one of the most fundamental properties
 of nonlinear dynamical systems.
It is often characterized by the Lyapunov exponents, i.e.,
 the exponential rates of divergence of infinitesimal perturbations
 given to a trajectory.
The Lyapunov exponents are also known to characterize
 properties of chaotic systems other than instability,
 such as the metric entropy and the attractor dimension \cite{Eckmann1985}.
Moreover, for large systems, the extensivity of chaos is defined
 on the basis of the spectrum of the Lyapunov exponents \cite{Ruelle1982}.
From the application point of view,
 the Lyapunov exponents and related objects play an important role
 in chaos control \cite{Andrievskii2004}
 and data assimilation \cite{Balci2012}.
It is therefore not surprising that the Lyapunov exponents have been
 central quantities to investigate in numerical studies of chaos,
 in which case the equation of motion is usually given
 and the methods to evaluate the exponents are established \cite{Pikovsky2016lyapunov,Shimada.Nagashima-PTP1979,Benettin.etal-M1980a}.
However, experimentally, the situation is in sharp contrast,
 because the equation of motion is usually unavailable and one often
 needs to resort to time series to estimate the Lyapunov exponents.

The most common experimental approach is the following
 \cite{Ott-Book2002,kantz2004nonlinear,Pikovsky2016lyapunov}:
(i) First, time series, say $s(t)$, are embedded
 to a space of sufficiently high dimensionality,
 by use of time-delayed coordinates
 $\bm{s}(t) = [s(t), s(t-\tau), s(t-2\tau), \cdots]$.
(ii) Recurrences of trajectories, i.e., pairs of $\bm{s}(t_i)$ and $\bm{s}(t_j)$
 with small $||\bm{s}(t_i) - \bm{s}(t_j)||$ are detected
 and the growth rate of $||\bm{s}(t_i+t) - \bm{s}(t_j+t)||$ is measured.
Although this method works well for systems
 with a small number of degrees of freedom,
 it cannot be applied to large systems whose number of degrees of freedom
 is large (typically $\gtrsim 10$), because recurrence becomes extremely rare
 in such high-dimensional space.
Recently, Pathak \textit{et al.} used a machine learning technique
 to time series data and succeeded in predicting trajectories
 and even Lyapunov exponents of a large spatially-extended system
 \cite{Pathak2017}.
This is an encouraging development,
 but adjusting many parameters involved in this method,
 without guiding principles, is presumably a delicate task in practice.

In this work, we choose to extend the recurrence method and attempt to overcome
 the problem of the lack of recurrences in large systems.
Here we restrict our target to a specific kind of systems, namely systems
 with global coupling, but we believe our method can be extended
 to other types of systems with a high degree of symmetry.
We focus on the fact that the evolution of a local variable does not necessarily
 require a large number of variables; in the case of globally coupled systems,
 it is determined only by the local variable and the mean field.
We therefore collect recurrences with this local set of variables
 and show that it is sufficient to construct the global Jacobian,
 which is necessary to compute the full spectrum of the Lyapunov exponents.
We apply our method to two globally-coupled systems, specifically,
 logistic maps and limit-cycle oscillators with global coupling,
 and demonstrate that this method is able to evaluate the Lyapunov spectrum
 reasonably well.
We also describe a few techniques
 to improve the accuracy of the proposed method.

\section{\label{sec:method}Method}

Here we describe our method for globally coupled systems in a general manner.
For simplicity, it is described for the case in which
 the local evolution is given by a one-dimensional map, but generalization
 to higher dimensions and to differential equations is straightforward.

Consider a globally coupled system given by
\begin{equation}
    x_j(t+1)
    =
    f( x_j(t), m(t) )
    \label{eq:globally_coupled_system}
\end{equation}
 with $j=1,2,\ldots,N$, where $f(x,m)$ is a nonlinear map,
 $x_j(t)$ represents the $j$th local variable
 at discrete time $t$, $m(t)$ is the mean field given by
\begin{equation}
 m(t):=\frac{1}{N} \sum_{j=1}^{N} x_{j}(t).  \label{eq:meanfield}
\end{equation}
The full Jacobian matrix for this system is
\begin{align}
  &J(x_{1}(t), \ldots, x_{N}(t)) \notag \\
    & \quad =
  \begin{bmatrix}
      \prt{}{x_1}f(x_{1}(t), m(t)), & \cdots, & \prt{}{x_N}f(x_{1}(t), m(t)) \\
        \vdots & & \vdots \\
        \prt{}{x_1}f(x_{N}(t), m(t)), & \cdots, & \prt{}{x_N}f(x_{N}(t), m(t)) \\
  \end{bmatrix}.  \label{eq:fullJacobian}
\end{align}
Here, note that the function $f(x,m)$
 takes two independent arguments $x$ and $m$,
 but since $m(t)$ is given by \eqref{eq:meanfield},
 the derivative in \eqref{eq:fullJacobian} should read
\begin{align}
 &\prt{}{x_i}f(x_j(t),m(t)) \notag \\
 &\qquad = \delta_{ij} \prt{f}{x}(x_j(t),m(t)) + \frac{1}{N}\prt{f}{m}(x_j(t),m(t))  \label{eq:Jacobianelement}
\end{align}
 with Kronecker's delta $\delta_{ij}$.
An important observation here is that the full Jacobian is determined only
 by the two derivatives of the local map, $\prt{f}{x}$ and $\prt{f}{m}$.
Therefore, time series data of a \textit{single} local variable $x_1(t)$
 and the mean field $m(t)$ are actually sufficient to reconstruct
 the full Jacobian matrix $J$.

We now describe the method.
Assume that we have time series of a single local variable $x_{1}(t)$
 and the mean field $m(t)$.
With $\bm{p}_{1}(t):=[x_{1}(t), m(t)]^{T}$,
 the total derivative of \eqref{eq:globally_coupled_system} is
\begin{equation}
  \mathrm{d} x_{1}(t+1)
    =
  \begin{bmatrix}
    A(\bm{p}_{1}(t)), & B(\bm{p}_{1}(t))
  \end{bmatrix}
    \mathrm{d}\bm{p}_{1}(t)
    \label{eq:generic_displacement_dynamics}
\end{equation}
where we define 
\begin{equation}
\begin{aligned}
 &A(\bm{p}_{1}(t)) := \prt{f}{x}(x_1(t),m(t)), \\
 &B(\bm{p}_{1}(t)) := \prt{f}{m}(x_1(t),m(t)). 
\end{aligned} \label{eq:AB}
\end{equation}
Note that \eqref{eq:generic_displacement_dynamics} is equivalent
 to the evolution of an infinitesimal perturbation $\rd x_1(t)$
 in a system with two $\textit{independent}$ variables $x_1(t)$ and $m(t)$,
 defined by \eqref{eq:globally_coupled_system}.
Therefore, we can use the standard recurrence method for this
 two-dimensional \textit{reduced} space spanned by $x_1(t)$ and $m(t)$,
 and obtain the $1 \times 2$ matrix $[ A(\bm{p}_{1}(t)), B(\bm{p}_{1}(t)) ]$
 which we shall call the pseudo local Jacobian matrix.
Specifically, adapting the method proposed
 by Sano and Sawada \cite{Sano.Sawada-PRL1985} and
 by Eckmann and Ruelle \cite{Eckmann1985,Eckmann.etal-PRA1986},
 we use pairs of recurrent points $\bm{p}_1(t_i)$ and $\bm{p}_1(t_j)$,
 regard $\mathrm{d}\bm{p}_{1}(t) \approx \bm{p}_1(t_i)-\bm{p}_1(t_j)$
 and $\mathrm{d} x_{1}(t+1) \approx x_1(t_i+1)-x_1(t_j+1)$,
 and evaluate the matrix $[ A(\bm{p}_{1}(t)), B(\bm{p}_{1}(t)) ]$
 by the least squares method.
Importantly, here we are able to obtain enough recurrences
 because the dimensionality of this reduced space
 is only two (or multiples of two
 if the local variable $x_j(t)$ is multidimensional).
Then, with \eqsref{eq:Jacobianelement} and \pref{eq:AB},
 we obtain the Jacobian matrix \pref{eq:fullJacobian} for the full system
 by appropriately interpolating $\prt{f}{x}(x,m)$ and $\prt{f}{m}(x,m)$.

To be precise, from time series data $\bm{p}_{1}(t) = [x_1(t), m(t)]^T$,
 we evaluate the Lyapunov exponents by the following two steps.
\textit{Step I}. 
We estimate the pseudo local Jacobian matrix
 $[ A(\bm{p}_{1}(t)), B(\bm{p}_{1}(t)) ]$
 by recurrences of time series $\bm{p}_{1}(t)$ in the reduced space.
To detect recurrences, we consider a small ball of radius $\epsilon$
 centered at a given $\bm{p}_{1}(t)$,
 and obtain the set of the indices of the recurrences,
 $I_\epsilon(t) := \{t'; ||\bm{p}_1(t)-\bm{p}_1(t')|| < \epsilon\}$.
Then, by the least squares method as described above, we obtain
 an estimate of the pseudo local Jacobian matrix,
 denoted by $[ \tilde{A}(\bm{p}_{1}(t)), \tilde{B}(\bm{p}_{1}(t)) ]$.
\textit{Step II}.
We numerically emulate
 both phase-space and tangent-space dynamics
 by interpolating the functions $f(x,m)$, $\prt{f}{x}(x,m)$,
 and $\prt{f}{m}(x,m)$,
 from time series data for $x_1(t+1)=f(x_1(t),m(t))$,
 $\tilde{A}(\bm{p}_{1}(t)) \approx \prt{f}{x}(x_1(t),m(t))$,
 and $\tilde{B}(\bm{p}_{1}(t)) \approx \prt{f}{m}(x_1(t),m(t))$,
 respectively.
Then the Lyapunov exponents are obtained
 by the standard QR decomposition method
 \cite{Pikovsky2016lyapunov,Shimada.Nagashima-PTP1979,Benettin.etal-M1980a}.

In the next two sections,
 we test our method with numerically generated time series,
 using globally coupled logistic maps (\secref{sec:GCM})
 and globally coupled limit-cycle oscillators
 (\secref{sec:limit_cycle_oscillator}).

\section{\label{sec:GCM}Globally coupled logistic maps}
\subsection{System}

We first consider a system of globally coupled logistic maps
\begin{equation}
\begin{gathered}
    x_j(t+1)
    =
    f( X_j(t) ),
    \\
    X_j(t) 
    = 
    (1-K) x_j(t) + K m(t),
    \label{eq:globally_coupled_map}
\end{gathered}
\end{equation}
 with $j=1,2,\ldots,N$, a coupling constant $K$,
 and the logistic map $f(x) = 1 - ax^2$.
Then the $1 \times 2$ pseudo local Jacobian matrix
 $[ A(\bm{p}_1(t)), B(\bm{p}_1(t)) ]$  is given by
\begin{align}
  \begin{split}
      A(\bm{p}_1(t))
      &=
      (1-K) f'(X_1(t)) ,
      \\
      B(\bm{p}_1(t))
      &=
      K f'(X_1(t)) .
    \end{split}  \label{eq:GCMAB}
\end{align}

In the following, we set $K=0.1$, $a=2$ and $N=200$.
This corresponds to a regime of high-dimensional chaos \cite{Kaneko1990b},
 which does not show any apparent coherence
 in the values of the local variables
 (in particular there is no synchronization at least in the usual sense).
We assume that we know the system to analyze has
 a global coupling in the additive form,
 as expressed generically by \eqref{eq:globally_coupled_map},
 but the function $f(x)$ is unknown.
We used time series data of a local variable $x_{1}(t)$
 and the mean field $m(t)$, generated numerically after discarding a transient.
The length of the time series data was $T=10^5$.
We applied our method described in \secref{sec:method}.
The radius of the $\epsilon$-ball neighborhood was set to $\epsilon=10^{-2}$.

\subsection{Results}

\begin{figure}[tp]
\includegraphics[width=\columnwidth]{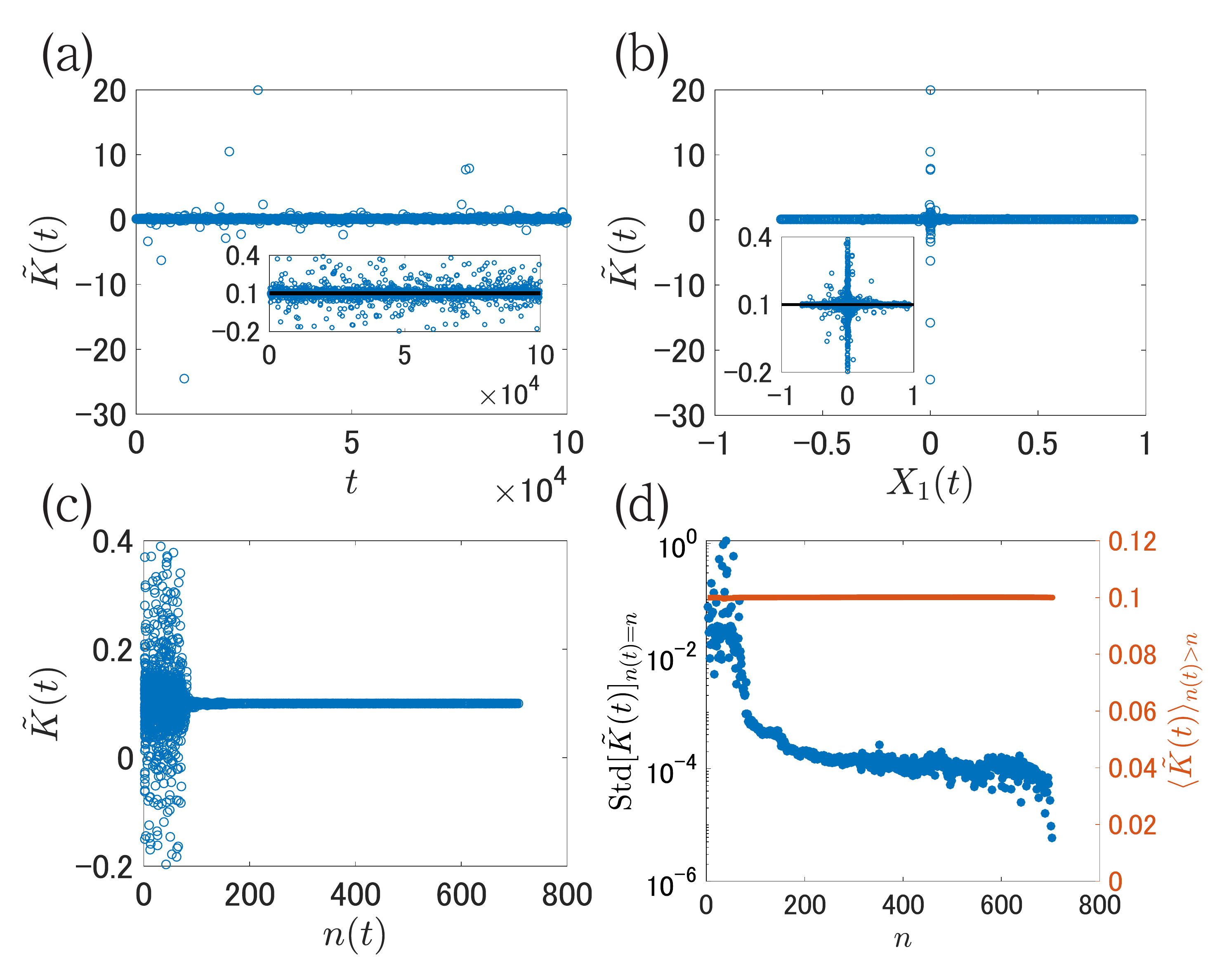}
\caption{
Estimation of the coupling constant $K$ for the globally coupled logistic maps.
(a,b) Estimates $\tilde{K}(t)$ shown against $t$ (a) and $X_1(t)$ (b).
The insets are close-ups showing the range $-0.4 \leq \tilde{K}(t) \leq 0.6$,
 with the true value $K=0.1$ indicated by the black solid line.
(c) Estimates $\tilde{K}(t)$ shown against the number of recurrences
 $n(t) :=|I_{\epsilon}(\bm{p}_1(t))|$.
(d) Standard deviation of the estimates $\tilde{K}(t)$
 with $n(t)=n$ (blue dots) and the mean of $\tilde{K}(t)$ such that $n(t)>n$
 (orange line) shown against $n$.
}
\label{fig:fig1}
\end{figure}

First, following Step I described in \secref{sec:method},
 we evaluate the coupling constant $K$.
Using \eqref{eq:globally_coupled_map}, we have
 $K = B(\bm{p}_1(t)) / (A(\bm{p}_1(t)) + B(\bm{p}_1(t)))$.
Therefore, from the estimates
 $[ \tilde{A}(\bm{p}_{1}(t)), \tilde{B}(\bm{p}_{1}(t)) ]$
 of the pseudo local Jacobian matrix, we obtain
\begin{equation}
 \tilde{K}(t) = \frac{ \tilde{B}(\bm{p}_1(t)) }{ \tilde{A}(\bm{p}_1(t)) + \tilde{B}(\bm{p}_1(t)) }.  \label{eq:GCM_K}
\end{equation}
Note that, though the true coupling parameter $K$ is a constant,
 it is evaluated for each data point $\bm{p}_1(t)$,
 so that $\tilde{K}(t)$ is a function of $t$.

\begin{figure}[!t]
\includegraphics[width=1.05\columnwidth]{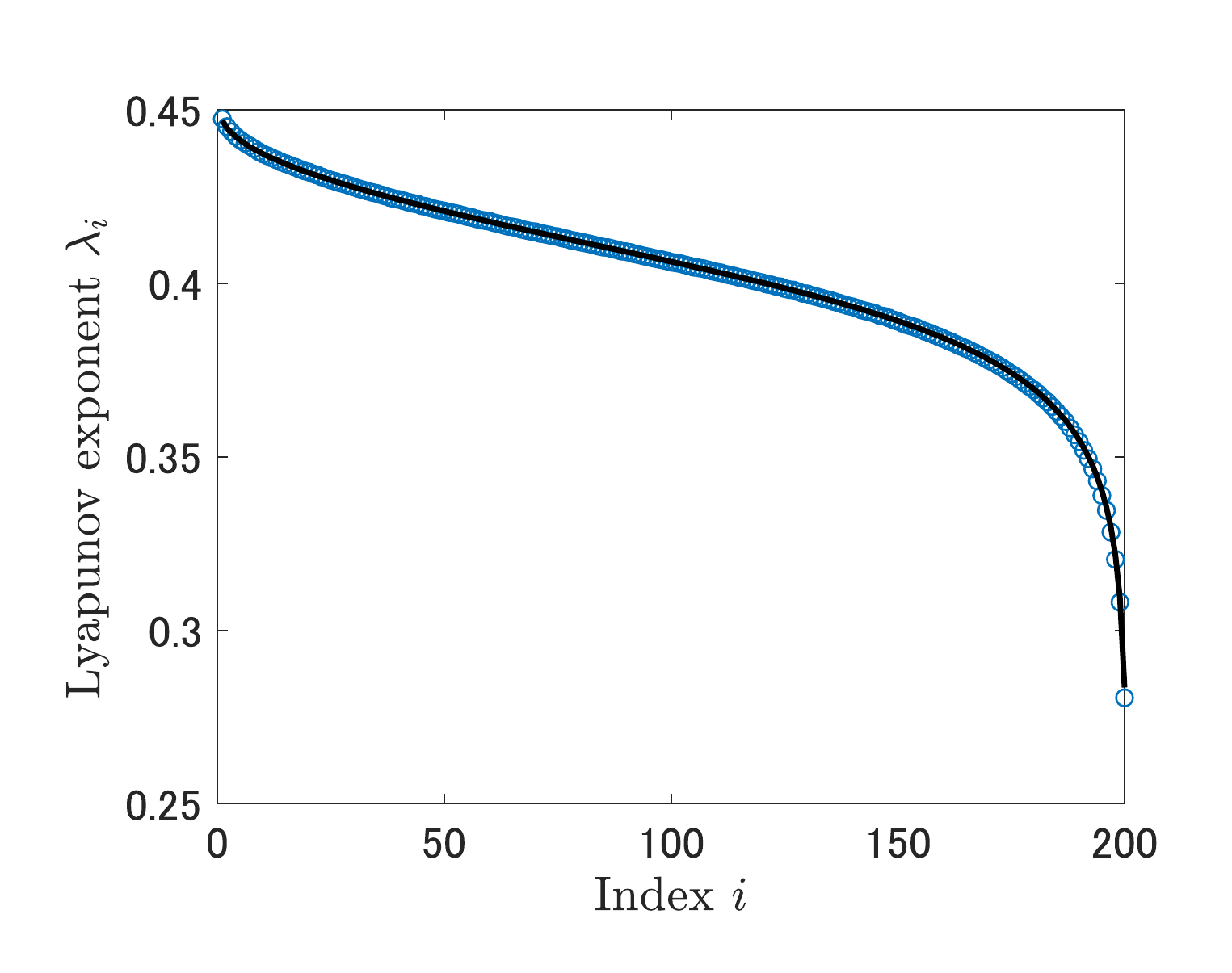}
\caption{
The spectrum of the Lyapunov exponents $\lambda_i$
 for the globally coupled logistic maps,
 evaluated by the proposed method (blue circles).
The black line indicates the true spectrum obtained
 by the standard QR decomposition method.
}
\label{fig:fig2}
\end{figure}

Figure~\ref{fig:fig1}(a) shows $\tilde{K}(t)$ as a function of time.
By taking the time average, we obtain $\langle \tilde{K}(t) \rangle \approx 0.09998$,
 which differs from the true value $K=0.1$ only by the order of $10^{-5}$.
However, the standard deviation of $\tilde{K}(t)$ is actually
 as large as $0.13$, which is also apparent from scattered data points
 in \figref{fig:fig1}(a).
 
A closer look reveals that errors are anomalously large
 when $X_1(t) \approx 0$ [\figref{fig:fig1}(b)].
This is easy to understand, because $f'(X_1(t)) = -2aX_1(t)$ is then
 almost vanishing and so is $\rd x_1(t+1)$
 given by \eqsref{eq:generic_displacement_dynamics} and \pref{eq:GCMAB}.
 
Therefore, the estimation
 of $[ \tilde{A}(\bm{p}_{1}(t)), \tilde{B}(\bm{p}_{1}(t)) ]$,
 or equivalently that of $\tilde{K}(t)$ and $\tilde{f}'(X_1(t))$,
 becomes numerically unstable for those particular data points.
The remaining source of error is the lack of recurrences.
In \figref{fig:fig1}(c), the estimates $\tilde{K}(t)$ are plotted against $n(t) :=|I_{\epsilon}(\bm{p}_1(t))|$,
 i.e., the number of the recurrence points around the data point $\bm{p}_1(t)$.
It is clear that large errors are essentially originated
 from data points with small $n$.
This is quantified in \figref{fig:fig1}(d),
 which shows how the standard deviation of $\tilde{K}(t)$
 with a given number of recurrences $n$,
 denoted by $\Std[\tilde{K}(t)]_{n(t)=n}$,
 decreases with increasing $n$ (blue dots).
We can see that the error level becomes very low, in the order of $10^{-4}$,
 for $n \gtrsim 100$.
Errors are not negligible for smaller $n$, but even so,
 the number of such data points is small enough
 so that the mean of $\tilde{K}(t)$ such that $n(t)>n$,
 denoted by $\expct{\tilde{K}(t)}_{n(t)>n}$, is hardly affected
 by the choice of the threshold $n$ [orange line in \figref{fig:fig1}(d)].

In any case, we obtain a reasonable estimate for the coupling constant,
 $\tilde{K} := \langle \tilde{K}(t) \rangle \approx 0.09998$.
The derivative $f'(X)$ is evaluated, from \eqref{eq:GCMAB}, by
 $\tilde{f}'(\tilde{X}_1(t)) = \tilde{A}(\bm{p}_1(t)) + \tilde{B}(\bm{p}_1(t))$
 with $\tilde{X}_1(t) := (1-K)x_1(t) + \tilde{K}m(t)$.
Then we carry out Step II in \secref{sec:method}
 and evaluate the Lyapunov exponents.
Figure~\ref{fig:fig2} shows the result (blue circles),
 compared with the true spectrum (black line) which we obtain directly
 by applying the QR decomposition method to the globally coupled logistic maps.
It is confirmed that our method successfully evaluated the Lyapunov exponents
 in the entire spectrum.

\section{\label{sec:limit_cycle_oscillator}Globally Coupled Limit Cycle Oscillator}

\subsection{System}

For the second example, we choose a system with continuous time,
 specifically a system of limit-cycle oscillators with global coupling,
 defined as follows:
\begin{widetext}
\begin{equation}
    \dot{w}_{j}(t)
    =
    w_{j}(t)
    -
    ( 1 + c_2)
    | w_{j}(t) |^2
    w_{j}(t)
    +
        K ( 1 + \textrm{i} c_1 )
    ( \bar{w}(t) - w_{j}(t) )
    \label{eq:limit_cycle_oscillator}
\end{equation}
 with $j=1,2,\ldots,N$, complex variables $w_{j}(t)$,
 the mean field $\bar{w}(t) := (1/N)\sum_j w_j(t)$,
 a coupling constant $K$, and system parameters $c_1,c_2$.
To write down the pseudo local Jacobian matrix, it is convenient to use
 $x_j(t) := \re [w_j(t)]$ and $y_j(t) := \im [w_j(t)]$,
 and discretize time by the Euler method with time step $\Delta t$.
The resulting submatrices $A(\bm{p}_{1}(t))$ and $B(\bm{p}_{1}(t))$,
 which are now $2 \times 2$
 with $\bm{p}_1(t) := [x_1(t), y_1(t), \bar{x}(t), \bar{y}(t)]^T$, read
\begin{equation}
\begin{aligned}
      &A(\bm{p}_{1}(t))
      =
      \begin{bmatrix}
        1 - 3 x_1^{2}(t) - y_1^{2}(t) + 2c_{2} x_1(t) y_1(t) - K, & c_{2} x_1^{2}(t) + 3 c_{2} y_1^{2}(t) - 2 x_1(t) y_1(t) + Kc_{1} \\
        - 3 c_{2} x_1^{2}(t) - c_{2} y_1^{2}(t) - 2 x_1(t) y_1(t) - Kc_{1}, & 1 - x_1^{2}(t) - 3 y_1^{2}(t) - 2c_{2} x_1(t) y_1(t) - K
      \end{bmatrix} \Delta t
      +
      \begin{bmatrix}
        1 & 0 \\
        0 & 1
      \end{bmatrix}
  \\
      &B(\bm{p}_{1}(t))
      =
      \begin{bmatrix}
        K & -Kc_{1} \\
        Kc_{1} & K
      \end{bmatrix} \Delta t
      .
\end{aligned}
\end{equation}

In the following, we set $K=0.52$, $c_1=-2.5$, $c_2=3.0$,
 which correspond to a regime of high-dimensional chaos \cite{Nakagawa1994,N.Nakagawa_PhysicaD_1995},
 and the system size is set to be $N=50$.
Again, the oscillators are not synchronized,
 but distributed in the complex plane \cite{Nakagawa1994,N.Nakagawa_PhysicaD_1995}.

For the analysis, we assume that we know the target is a system described in the following form:
\begin{equation}
\begin{aligned}
 &\dot{x}_j(t) = f_x(x_j(t),y_j(t))
 + K_{xx} (\bar{x}(t)-x_j(t)) + K_{xy} (\bar{y}(t)-y_j(t)),  \\
 &\dot{y}_j(t) = f_y(x_j(t),y_j(t))
 + K_{yx} (\bar{x}(t)-x_j(t)) + K_{yy} (\bar{y}(t)-y_j(t)), 
\end{aligned}  \label{eq:LMO_assumption}
\end{equation}
 but the functional forms of $f_x(x,y)$ and $f_y(x,y)$,
 as well as the values of the four coupling constants are unknown.
The pseudo local Jacobian matrix then reads:
\begin{equation}
\begin{aligned}
 &A(x_1,y_1) = \begin{bmatrix}
 \prt{f_x}{x}(x_1,y_1)-K_{xx}, & \prt{f_x}{y}(x_1,y_1)-K_{xy} \\
 \prt{f_y}{x}(x_1,y_1)-K_{yx}, & \prt{f_y}{y}(x_1,y_1)-K_{yy}
 \end{bmatrix} \Delta t
 + \begin{bmatrix} 1 & 0 \\ 0 & 1 \end{bmatrix}, \\
 &B =
 \begin{bmatrix} K_{xx} & K_{xy} \\ K_{yx} & K_{yy} \end{bmatrix} \Delta t.
\end{aligned}  \label{eq:LMO_assumption2}
\end{equation}
\end{widetext}
Note that, thanks to the linear coupling to the mean field,
 the matrix $A(\bm{p}_1)$ depends only on $x_1$ and $y_1$,
 and $B(\bm{p}_1)$ is a constant matrix.

We used time series of a local variable $w_1(t) = x_1(t) + \ri y_1(t)$
 and the mean field $\bar{w}(t) = \bar{x}(t) + \ri \bar{y}(t)$,
 generated numerically by the fourth-order Runge-Kutta method
 with time step $\Delta t = 10^{-3}$, after discarding a transient.
The length of the time series data was $T=10^6$ (in the unit of time step).
Then we applied our method with $\epsilon=10^{-2}$
 and evaluated the coupling constants and the Lyapunov exponents.

\subsection{Results}

\begin{figure}[htp]
\includegraphics[width=\columnwidth]{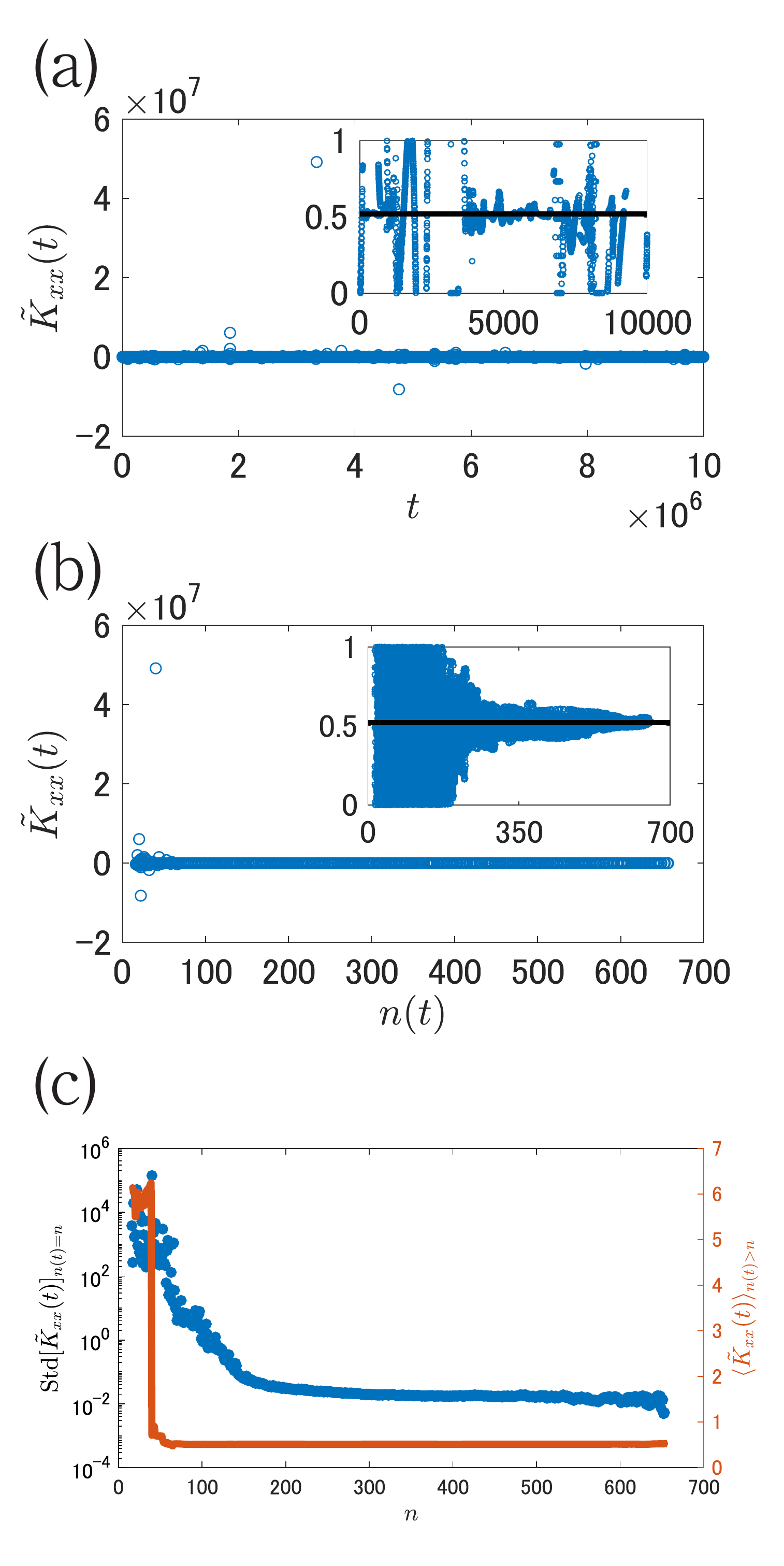}
\caption{
Estimation of the coupling constant $K_{xx}$
 for the globally coupled limit-cycle oscillators.
(a,b) Estimates $\tilde{K}_{xx}(t)$ shown against time $t$ (a)
 and the number of recurrences $n(t) :=|I_{\epsilon}(\bm{p}_1(t))|$ (b).
The insets are close-ups showing the range $0 \leq \tilde{K}_{xx}(t) \leq 1$,
 with the true value $K=0.52$ indicated by the black solid line.
(c) Standard deviation of the estimates $\tilde{K}_{xx}(t)$
 with $n(t)=n$ (blue dots) and the mean of $\tilde{K}_{xx}(t)$
 such that $n(t)>n$ (orange line) shown against $n$.
}
\label{fig:fig3}
\end{figure}

Similarly to the procedure we adopted in \secref{sec:GCM}, by Step I,
 we first evaluate the coupling constants.
Taking $K_{xx}$ as an example, from \eqref{eq:LMO_assumption2}
 we obtain $\tilde{K}_{xx}(t) = \tilde{B}(\bm{p}_{1}(t))/\Delta t$
 [\figref{fig:fig3}(a)].
The data suggest that, compared to the previous case, the estimates
 $\tilde{K}_{xx}(t)$ tend to meander far from the true value
 $K_{xx} = K = 0.52$ for longer time.
Indeed, simple time averaging now yields a totally wrong value,
 $\expct{\tilde{K}_{xx}(t)} \approx 6.08$.
On the other hand, we find that the median gives a reasonable value $0.515$,
 suggesting that $\tilde{K}_{xx}(t)$ still spends
 much time near the true value
 [see also the inset of \figref{fig:fig3}(a)].

The estimation accuracy can be improved
 by paying attention to the number of recurrences.
Figures~\ref{fig:fig3}(b) and (c) display $\tilde{K}_{xx}(t)$
 against $n(t) = |I_\epsilon(\bm{p}_1(t))|$ [panel (b)],
 as well as $\Std[\tilde{K}_{xx}(t)]_{n(t)=n}$
 [blue dots of panel (c)] and
 $\expct{\tilde{K}_{xx}(t)}_{n(t)>n}$ (orange line) against $n$.
These results consistently show that most errors in $\expct{\tilde{K}_{xx}(t)}$
 are due to the data points with only few recurrent points.
Therefore, we can improve the accuracy by setting a lower threshold for $n$,
 denoted by $n_\textrm{trm}$, and using only the data points
 with $n(t)>n_\textrm{trm}$.
We shall call this operation ``trimming'', 
 and $n_\textrm{trm}$ the trimming threshold.
Figure~\ref{fig:fig3}(c) shows that, with $n_\textrm{trm} \approx 50$,
 the mean estimate $\expct{\tilde{K}_{xx}(t)}_{n(t)>n}$ is already stable
 (orange line)
 but individual estimates $\tilde{K}_{xx}(t)$ are still fluctuating
 (blue dots).
The fluctuation level becomes low for $n \gtrsim 200$ or $300$
 [see also \figref{fig:fig3}(b)], so that these are expected to be
 an appropriate choice for the value of $n_\textrm{trm}$.

\begin{figure}[htp]
\includegraphics[width=\columnwidth]{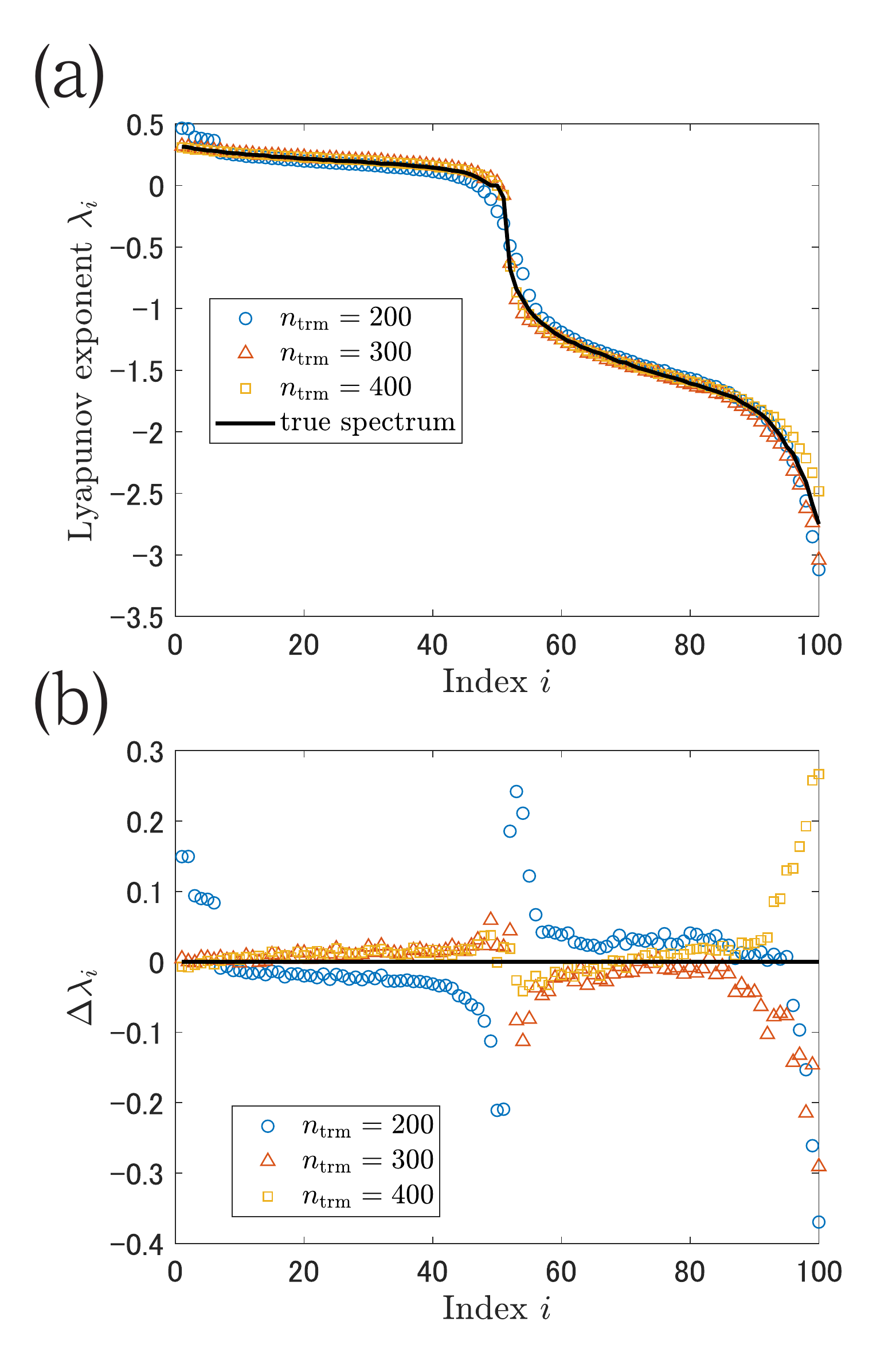}
\caption{
The spectrum of the Lyapunov exponents $\lambda_i$
 for the globally coupled limit-cycle oscillators.
(a) The spectrum evaluated by the proposed method (symbols)
 for different choices of the trimming threshold $n_\textrm{trm}$.
The black line indicates the true spectrum $\lambda_i^\textrm{true}$ obtained
 by the standard QR decomposition method.
(b) The estimation error
 $\Delta\lambda_i := \lambda_i - \lambda_i^\textrm{true}$.
}
\label{fig:fig4}
\end{figure}

Now we evaluate the Lyapunov exponents via Step II,
 i.e., by emulating the phase-space and tangent-space dynamics.
The phase-space dynamics is realized by the time evolution equation
 \pref{eq:LMO_assumption}. Here, for the coupling constants
 the values obtained previously with the trimming technique are used,
 and the functions $f_x(x_j,y_j)$ and $f_y(x_j,y_j)$ are evaluated
 by interpolation of the time series data.
The tangent-space dynamics is reconstructed
 by interpolating the estimates of the matrix $A(x_j,y_j)$
 [\eqref{eq:LMO_assumption2}], while for $B$ the obtained values
 of the coupling constants are used.
For those interpolations, we again need to have sufficiently many
 neighbors around each time-series data point.
In fact, we can increase this number in the case where
 we know \textit{a priori} that our oscillators are invariant
 under uniform shift of the phase, i.e.,
 under the transformation $w_{j}(t) \to w_{j}(t)e^{i \theta}$
 with a constant $\theta$ for all $j$.
Specifically, if we are to evaluate $A(x_j,y_j)$, or equivalently $A(w_j)$,
 we only need to find $w_1(t)$ from the time series data such that
 the modulus $|w_1(t)|$ is close to $|w_j|$.
Then we rotate $w_1(t)$ by the angle $\theta = \arg w_j - \arg w_1(t)$,
 or, more precisely, transform $A(w_1(t))$ to
 $R(\theta) A(w_{1}(t)) R^{-1}(\theta)$ with the rotation matrix
 $R(\theta) := \begin{bmatrix} \cos\theta & -\sin\theta \\ \sin\theta & \cos\theta \end{bmatrix}$,
 and interpolate the value on the one-dimensional number line.
The interpolation of $f_x(x,y)$ and $f_y(x,y)$ can also be done analogously,
 for which we use the fourth-order central-difference formula
 to evaluate $\dot{w}_1$ from the time series $w_1(t)$.

Figure~\ref{fig:fig4}(a) shows the Lyapunov spectrum obtained by our method
 (symbols), with varying trimming threshold $n_\mathrm{trm}$,
 compared with the true spectrum (black line), which is obtained by
 using the QR decomposition method
 to the limit-cycle oscillators \pref{eq:limit_cycle_oscillator}.
The difference from the true spectrum is displayed in \figref{fig:fig4}(b).
We can confirm that our results reproduce the true spectrum reasonably well.

\section{\label{sec:conclusion}Conclusions}

In this work we proposed a method to evaluate the Lyapunov exponents
 from time series data of large chaotic systems with global coupling.
The central idea is to handle the recurrence analysis
 in the reduced space,
 which consists only of a local variable and the mean field,
 thus circumventing the usual difficulty of the lack of recurrence points.
We demonstrated the validity of our method with two representative systems,
 namely the globally coupled logistic maps and the globally coupled limit-cycle
 oscillators, and reproduced the true Lyapunov spectrum reasonably well.
It is true that systems with global coupling, which we consider in this work, are a specific kind of large dynamical systems.
However, there are real examples of such systems, as shown by laboratory experiments of chaotic electrochemical oscillators\cite{Wang2000} and metabolic oscillations of stirred yeast cells \cite{D.Monte_PNAS_2007}.
In general, well-mixed many-component systems can often be regarded
 as systems with global coupling.
Those systems are potential targets for applying our method experimentally.

Compared to the recently proposed method
 based on the machine learning technique \cite{Pathak2017},
 which does not require \textit{a priori} assumptions on the form of coupling,
 the advantage of our method is that the adjustable parameters are much fewer:
 specifically, the cutoff $\epsilon$ for the detection of recurrences
 and the trimming threshold $n_\mathrm{trm}$,
 whose physical meaning is also clear.
Our method can also be extended to other types of systems
 that have a high degree of symmetry, in the sense that
 the evolution of a local dynamical variable is determined
 by a small number of variables.
We are aware that, for applying our method to experimental systems,
 we also need to incorporate the embedding technique
 \cite{Ott-Book2002,kantz2004nonlinear,Pikovsky2016lyapunov},
 as well as to evaluate the influence of noise and inhomogeneity
 -- important tasks left for future studies.
We believe that the results presented here make the first step
 on this track, towards the realization of instability analysis
 of large experimental systems.

\begin{acknowledgments}
We would like to thank R. Tosaka for useful discussions.
This work is supported in part by KAKENHI
 from Japan Society for the Promotion of Science
 (No. JP16K13846, JP16H04033, JP25103004).
\end{acknowledgments}

\bibliography{for_my_paper,book,ref-add}

\begin{thebibliography}{19}%
\makeatletter
\providecommand \@ifxundefined [1]{%
 \@ifx{#1\undefined}
}%
\providecommand \@ifnum [1]{%
 \ifnum #1\expandafter \@firstoftwo
 \else \expandafter \@secondoftwo
 \fi
}%
\providecommand \@ifx [1]{%
 \ifx #1\expandafter \@firstoftwo
 \else \expandafter \@secondoftwo
 \fi
}%
\providecommand \natexlab [1]{#1}%
\providecommand \enquote  [1]{``#1''}%
\providecommand \bibnamefont  [1]{#1}%
\providecommand \bibfnamefont [1]{#1}%
\providecommand \citenamefont [1]{#1}%
\providecommand \href@noop [0]{\@secondoftwo}%
\providecommand \href [0]{\begingroup \@sanitize@url \@href}%
\providecommand \@href[1]{\@@startlink{#1}\@@href}%
\providecommand \@@href[1]{\endgroup#1\@@endlink}%
\providecommand \@sanitize@url [0]{\catcode `\\12\catcode `\$12\catcode
  `\&12\catcode `\#12\catcode `\^12\catcode `\_12\catcode `\%12\relax}%
\providecommand \@@startlink[1]{}%
\providecommand \@@endlink[0]{}%
\providecommand \url  [0]{\begingroup\@sanitize@url \@url }%
\providecommand \@url [1]{\endgroup\@href {#1}{\urlprefix }}%
\providecommand \urlprefix  [0]{URL }%
\providecommand \Eprint [0]{\href }%
\providecommand \doibase [0]{http://dx.doi.org/}%
\providecommand \selectlanguage [0]{\@gobble}%
\providecommand \bibinfo  [0]{\@secondoftwo}%
\providecommand \bibfield  [0]{\@secondoftwo}%
\providecommand \translation [1]{[#1]}%
\providecommand \BibitemOpen [0]{}%
\providecommand \bibitemStop [0]{}%
\providecommand \bibitemNoStop [0]{.\EOS\space}%
\providecommand \EOS [0]{\spacefactor3000\relax}%
\providecommand \BibitemShut  [1]{\csname bibitem#1\endcsname}%
\let\auto@bib@innerbib\@empty
\bibitem [{\citenamefont {Frisch}(1995)}]{frisch1995turbulence}%
  \BibitemOpen
  \bibfield  {author} {\bibinfo {author} {\bibfnamefont {U.}~\bibnamefont
  {Frisch}},\ }\href@noop {} {\emph {\bibinfo {title} {Turbulence: the legacy
  of A. N. Kolmogorov}}}\ (\bibinfo  {publisher} {Cambridge Univ. Press},\
  \bibinfo {year} {1995})\BibitemShut {NoStop}%
\bibitem [{\citenamefont {Wang}, \citenamefont {Kiss},\ and\ \citenamefont
  {Hudson}(2000)}]{Wang2000}%
  \BibitemOpen
  \bibfield  {author} {\bibinfo {author} {\bibfnamefont {W.}~\bibnamefont
  {Wang}}, \bibinfo {author} {\bibfnamefont {I.~Z.}\ \bibnamefont {Kiss}}, \
  and\ \bibinfo {author} {\bibfnamefont {J.}~\bibnamefont {Hudson}},\
  }\href@noop {} {\bibfield  {journal} {\bibinfo  {journal} {Chaos}\ }\textbf
  {\bibinfo {volume} {10}},\ \bibinfo {pages} {248} (\bibinfo {year}
  {2000})}\BibitemShut {NoStop}%
\bibitem [{\citenamefont {Cherry}\ and\ \citenamefont
  {Fenton}(2008)}]{Alonso2008}%
  \BibitemOpen
  \bibfield  {author} {\bibinfo {author} {\bibfnamefont {E.~M.}\ \bibnamefont
  {Cherry}}\ and\ \bibinfo {author} {\bibfnamefont {F.~H.}\ \bibnamefont
  {Fenton}},\ }\href@noop {} {\bibfield  {journal} {\bibinfo  {journal} {New J.
  Phys.}\ }\textbf {\bibinfo {volume} {10}},\ \bibinfo {pages} {125016}
  (\bibinfo {year} {2008})}\BibitemShut {NoStop}%
\bibitem [{\citenamefont {Ott}(2002)}]{Ott-Book2002}%
  \BibitemOpen
  \bibfield  {author} {\bibinfo {author} {\bibfnamefont {E.}~\bibnamefont
  {Ott}},\ }\href@noop {} {\emph {\bibinfo {title} {Chaos in Dynamical
  Systems}}},\ \bibinfo {edition} {2nd}\ ed.\ (\bibinfo  {publisher} {Cambridge
  Univ. Press},\ \bibinfo {address} {Cambridge},\ \bibinfo {year}
  {2002})\BibitemShut {NoStop}%
\bibitem [{\citenamefont {Eckmann}\ and\ \citenamefont
  {Ruelle}(1985)}]{Eckmann1985}%
  \BibitemOpen
  \bibfield  {author} {\bibinfo {author} {\bibfnamefont {J.-P.}\ \bibnamefont
  {Eckmann}}\ and\ \bibinfo {author} {\bibfnamefont {D.}~\bibnamefont
  {Ruelle}},\ }\href@noop {} {\bibfield  {journal} {\bibinfo  {journal} {Rev.
  Mod. Phys.}\ }\textbf {\bibinfo {volume} {57}},\ \bibinfo {pages} {617}
  (\bibinfo {year} {1985})}\BibitemShut {NoStop}%
\bibitem [{\citenamefont {Ruelle}(1982)}]{Ruelle1982}%
  \BibitemOpen
  \bibfield  {author} {\bibinfo {author} {\bibfnamefont {D.}~\bibnamefont
  {Ruelle}},\ }\href@noop {} {\bibfield  {journal} {\bibinfo  {journal}
  {Commun. Math. Phys.}\ }\textbf {\bibinfo {volume} {87}},\ \bibinfo {pages}
  {287} (\bibinfo {year} {1982})}\BibitemShut {NoStop}%
\bibitem [{\citenamefont {Andrievskii}\ and\ \citenamefont
  {Fradkov}(2004)}]{Andrievskii2004}%
  \BibitemOpen
  \bibfield  {author} {\bibinfo {author} {\bibfnamefont {B.}~\bibnamefont
  {Andrievskii}}\ and\ \bibinfo {author} {\bibfnamefont {A.}~\bibnamefont
  {Fradkov}},\ }\href@noop {} {\bibfield  {journal} {\bibinfo  {journal}
  {Autom. Remote Control}\ }\textbf {\bibinfo {volume} {65}},\ \bibinfo {pages}
  {505} (\bibinfo {year} {2004})}\BibitemShut {NoStop}%
\bibitem [{\citenamefont {Balci}\ \emph {et~al.}(2012)\citenamefont {Balci},
  \citenamefont {Mazzucato}, \citenamefont {Restrepo},\ and\ \citenamefont
  {Sell}}]{Balci2012}%
  \BibitemOpen
  \bibfield  {author} {\bibinfo {author} {\bibfnamefont {N.}~\bibnamefont
  {Balci}}, \bibinfo {author} {\bibfnamefont {A.~L.}\ \bibnamefont
  {Mazzucato}}, \bibinfo {author} {\bibfnamefont {J.~M.}\ \bibnamefont
  {Restrepo}}, \ and\ \bibinfo {author} {\bibfnamefont {G.~R.}\ \bibnamefont
  {Sell}},\ }\href@noop {} {\bibfield  {journal} {\bibinfo  {journal} {Mon.
  Weather Rev.}\ }\textbf {\bibinfo {volume} {140}},\ \bibinfo {pages} {2308}
  (\bibinfo {year} {2012})}\BibitemShut {NoStop}%
\bibitem [{\citenamefont {Pikovsky}\ and\ \citenamefont
  {Politi}(2016)}]{Pikovsky2016lyapunov}%
  \BibitemOpen
  \bibfield  {author} {\bibinfo {author} {\bibfnamefont {A.}~\bibnamefont
  {Pikovsky}}\ and\ \bibinfo {author} {\bibfnamefont {A.}~\bibnamefont
  {Politi}},\ }\href@noop {} {\emph {\bibinfo {title} {Lyapunov exponents: a
  tool to explore complex dynamics}}}\ (\bibinfo  {publisher} {Cambridge Univ.
  Press},\ \bibinfo {year} {2016})\BibitemShut {NoStop}%
\bibitem [{\citenamefont {Shimada}\ and\ \citenamefont
  {Nagashima}(1979)}]{Shimada.Nagashima-PTP1979}%
  \BibitemOpen
  \bibfield  {author} {\bibinfo {author} {\bibfnamefont {I.}~\bibnamefont
  {Shimada}}\ and\ \bibinfo {author} {\bibfnamefont {T.}~\bibnamefont
  {Nagashima}},\ }\href@noop {} {\bibfield  {journal} {\bibinfo  {journal}
  {Prog. Theor. Phys.}\ }\textbf {\bibinfo {volume} {61}},\ \bibinfo {pages}
  {1605} (\bibinfo {year} {1979})}\BibitemShut {NoStop}%
\bibitem [{\citenamefont {Benettin}\ \emph {et~al.}(1980)\citenamefont
  {Benettin}, \citenamefont {Galgani}, \citenamefont {Giorgilli},\ and\
  \citenamefont {Strelcyn}}]{Benettin.etal-M1980a}%
  \BibitemOpen
  \bibfield  {author} {\bibinfo {author} {\bibfnamefont {G.}~\bibnamefont
  {Benettin}}, \bibinfo {author} {\bibfnamefont {L.}~\bibnamefont {Galgani}},
  \bibinfo {author} {\bibfnamefont {A.}~\bibnamefont {Giorgilli}}, \ and\
  \bibinfo {author} {\bibfnamefont {J.-M.}\ \bibnamefont {Strelcyn}},\
  }\href@noop {} {\bibfield  {journal} {\bibinfo  {journal} {Meccanica}\
  }\textbf {\bibinfo {volume} {15}},\ \bibinfo {pages} {9} (\bibinfo {year}
  {1980})}\BibitemShut {NoStop}%
\bibitem [{\citenamefont {Kantz}\ and\ \citenamefont
  {Schreiber}(2004)}]{kantz2004nonlinear}%
  \BibitemOpen
  \bibfield  {author} {\bibinfo {author} {\bibfnamefont {H.}~\bibnamefont
  {Kantz}}\ and\ \bibinfo {author} {\bibfnamefont {T.}~\bibnamefont
  {Schreiber}},\ }\href@noop {} {\emph {\bibinfo {title} {Nonlinear time series
  analysis}}},\ Vol.~\bibinfo {volume} {7}\ (\bibinfo  {publisher} {Cambridge
  Univ. Press},\ \bibinfo {year} {2004})\BibitemShut {NoStop}%
\bibitem [{\citenamefont {Pathak}\ \emph {et~al.}(2017)\citenamefont {Pathak},
  \citenamefont {Lu}, \citenamefont {Hunt}, \citenamefont {Girvan},\ and\
  \citenamefont {Ott}}]{Pathak2017}%
  \BibitemOpen
  \bibfield  {author} {\bibinfo {author} {\bibfnamefont {J.}~\bibnamefont
  {Pathak}}, \bibinfo {author} {\bibfnamefont {Z.}~\bibnamefont {Lu}}, \bibinfo
  {author} {\bibfnamefont {B.~R.}\ \bibnamefont {Hunt}}, \bibinfo {author}
  {\bibfnamefont {M.}~\bibnamefont {Girvan}}, \ and\ \bibinfo {author}
  {\bibfnamefont {E.}~\bibnamefont {Ott}},\ }\href@noop {} {\bibfield
  {journal} {\bibinfo  {journal} {Chaos}\ }\textbf {\bibinfo {volume} {27}},\
  \bibinfo {pages} {121102} (\bibinfo {year} {2017})}\BibitemShut {NoStop}%
\bibitem [{\citenamefont {Sano}\ and\ \citenamefont
  {Sawada}(1985)}]{Sano.Sawada-PRL1985}%
  \BibitemOpen
  \bibfield  {author} {\bibinfo {author} {\bibfnamefont {M.}~\bibnamefont
  {Sano}}\ and\ \bibinfo {author} {\bibfnamefont {Y.}~\bibnamefont {Sawada}},\
  }\href@noop {} {\bibfield  {journal} {\bibinfo  {journal} {Phys. Rev. Lett.}\
  }\textbf {\bibinfo {volume} {55}},\ \bibinfo {pages} {1082} (\bibinfo {year}
  {1985})}\BibitemShut {NoStop}%
\bibitem [{\citenamefont {Eckmann}\ \emph {et~al.}(1986)\citenamefont
  {Eckmann}, \citenamefont {Kamphorst}, \citenamefont {Ruelle},\ and\
  \citenamefont {Ciliberto}}]{Eckmann.etal-PRA1986}%
  \BibitemOpen
  \bibfield  {author} {\bibinfo {author} {\bibfnamefont {J.-P.}\ \bibnamefont
  {Eckmann}}, \bibinfo {author} {\bibfnamefont {S.~O.}\ \bibnamefont
  {Kamphorst}}, \bibinfo {author} {\bibfnamefont {D.}~\bibnamefont {Ruelle}}, \
  and\ \bibinfo {author} {\bibfnamefont {S.}~\bibnamefont {Ciliberto}},\
  }\href@noop {} {\bibfield  {journal} {\bibinfo  {journal} {Phys. Rev. A}\
  }\textbf {\bibinfo {volume} {34}},\ \bibinfo {pages} {4971} (\bibinfo {year}
  {1986})}\BibitemShut {NoStop}%
\bibitem [{\citenamefont {Kaneko}(1990)}]{Kaneko1990b}%
  \BibitemOpen
  \bibfield  {author} {\bibinfo {author} {\bibfnamefont {K.}~\bibnamefont
  {Kaneko}},\ }\href@noop {} {\bibfield  {journal} {\bibinfo  {journal}
  {Physica D}\ }\textbf {\bibinfo {volume} {41}},\ \bibinfo {pages} {137}
  (\bibinfo {year} {1990})}\BibitemShut {NoStop}%
\bibitem [{\citenamefont {Nakagawa}\ and\ \citenamefont
  {Kuramoto}(1994)}]{Nakagawa1994}%
  \BibitemOpen
  \bibfield  {author} {\bibinfo {author} {\bibfnamefont {N.}~\bibnamefont
  {Nakagawa}}\ and\ \bibinfo {author} {\bibfnamefont {Y.}~\bibnamefont
  {Kuramoto}},\ }\href@noop {} {\bibfield  {journal} {\bibinfo  {journal}
  {Physica D}\ }\textbf {\bibinfo {volume} {75}},\ \bibinfo {pages} {74}
  (\bibinfo {year} {1994})}\BibitemShut {NoStop}%
\bibitem [{\citenamefont {Nakagawa}\ and\ \citenamefont
  {Kuramoto}(1995)}]{N.Nakagawa_PhysicaD_1995}%
  \BibitemOpen
  \bibfield  {author} {\bibinfo {author} {\bibfnamefont {N.}~\bibnamefont
  {Nakagawa}}\ and\ \bibinfo {author} {\bibfnamefont {Y.}~\bibnamefont
  {Kuramoto}},\ }\href@noop {} {\bibfield  {journal} {\bibinfo  {journal}
  {Physica D}\ }\textbf {\bibinfo {volume} {80}},\ \bibinfo {pages} {307}
  (\bibinfo {year} {1995})}\BibitemShut {NoStop}%
\bibitem [{\citenamefont {De~Monte}\ \emph {et~al.}(2007)\citenamefont
  {De~Monte}, \citenamefont {d'Ovidio}, \citenamefont {Dan{\o}},\ and\
  \citenamefont {S{\o}rensen}}]{D.Monte_PNAS_2007}%
  \BibitemOpen
  \bibfield  {author} {\bibinfo {author} {\bibfnamefont {S.}~\bibnamefont
  {De~Monte}}, \bibinfo {author} {\bibfnamefont {F.}~\bibnamefont {d'Ovidio}},
  \bibinfo {author} {\bibfnamefont {S.}~\bibnamefont {Dan{\o}}}, \ and\
  \bibinfo {author} {\bibfnamefont {P.~G.}\ \bibnamefont {S{\o}rensen}},\
  }\href@noop {} {\bibfield  {journal} {\bibinfo  {journal} {Proc. Natl. Acad.
  Sci. U.S.A.}\ }\textbf {\bibinfo {volume} {104}},\ \bibinfo {pages} {18377}
  (\bibinfo {year} {2007})}\BibitemShut {NoStop}%
\end{thebibliography}%

\end{document}